\documentclass{mem}
\usepackage{natbib}\usepackage{txfonts}\usepackage{balance}
\usepackage{flushend}
\usepackage{graphicx}
\usepackage[breaklinks,pdftex]{hyperref}
\begin{document}
\def\teff{$T\rm_{eff }$}
\def\kms{$\mathrm {km s}^{-1}$}

\def\psr{PSR J1509--5859}
\def\lsi{LS I +61$^{\circ}$ 303}

\title{
Galactic Science
}

   \subtitle{Rapporteur Talk of the  8$^{th}$
Heidelberg International Symposium on High Energy Gamma Ray Astronomy}

\author{
Sandro Mereghetti
}

\institute{
INAF, IASF-Milano
\email{sandro.mereghetti@inaf.it}\\
}

\authorrunning{S. Mereghetti}

\titlerunning{Galactic Science}

\date{Received: 15-02-2025; Accepted:  02-06-2025}

\abstract{
The most recent observational and theoretical results in the rapidly expanding field of high-energy gamma-ray astrophysics were discussed  at the international conference ``Gamma-2024''  that took place in Milano in September 2024.
This contribution summarises the 'rapporteur talk' relative to  the Galactic science given at the end of the conference.
}
\maketitle{}

\section{Introduction}
 
The   Gamma-2024 conference,  held on 2024 September 2-6,
gathered in Milano  more than two hundreds scientists from twenty-five  different nations  
to present and discuss the most recent advances in  the field of 
 VHE (0.1-100 TeV)  and UHE (0.1-100 PeV) astronomy.  
 This conference was the 8$^{th}$ meeting in the series of Heidelberg International Symposia on High Energy Gamma Ray Astronomy.

Ground-based observatories, Imaging Air Cherenkov Telescopes and Extensive Air Shower arrays, are providing first-class results,  with  major  discoveries announced every month.  Space-  plus ground-based $\gamma$-ray astronomy now extends over  more than ten decades of the electromagnetic spectrum,  an energy range almost as wide  as that covered from radio to X-ray wavelengths.  
 
The science community involved  with  VHE and UHE astrophysics has grown significantly not only because observations at these energies allow us to tackle fundamental questions, such as the  origin of cosmic rays or the  extreme properties of compact objects,   but also because  high-energy emission  has now been  detected in many different kinds of  sources.  
While in the exploratory era of high-energy astronomy we had to rely only on the data of the few brightest emitters (not always representatives of the typical members of their classes), the current facilities can  detect point like and extended sources with much fainter fluxes,  thus providing significant samples sources.  

This paper is  a brief overview of the Galactic science topics discussed at the conference, without attempting to  be a complete review 
nor a discussion of the  exciting prospects expected from future facilities, for which the reader can refer to other specific contributions in these proceedings.

 \section{Pulsars}
\label{sec:psr}

Rotation-powered neutron stars were among the first discovered  $\gamma$-ray sources, with both radio-loud (Vela and Crab) and radio-quiet pulsars (Geminga) prominent among the SAS-2 sources. Before the advent of the AGILE and Fermi satellites, only young and middle-aged pulsars with high spin-down luminosity were detected in  $\gamma$-rays, but it is now clear that the old millisecond pulsars constitute a much more numerous class of  MeV-GeV sources. In fact they account for about 50\% of the pulsars in the 3$^{rd}$ Fermi/LAT pulsar catalog \citep{2023ApJ...958..191S}.  

For many years, models for $\gamma$-ray   emission from pulsars considered  different acceleration regions in the magnetosphere, where the  electric field component parallel to the magnetic field lines is unscreened, either close to the polar caps or farther out,  up to  the light cylinder radius (outer gap models). Detailed predictions on pulsar spectra and pulse profiles were compared to the large Fermi/LAT database in order to constrain these models (see, e.g.  \citep{iniguezTP}).  Such studies, as well as the results of population syntheses simulations, led to favour outer gap models, with curvature radiation from electrons and positrons accelerated along the magnetic lines as the main contributor in the $\sim$0.1-10 GeV range. 
The subsequent development of global magnetosphere models has significantly changed our perspective on pulsar emission, showing that particle  acceleration occurs mostly beyond the light cylinder.   Possible sites are the separatrix  and current sheet regions in the striped wind   \citep{hardingTP}. 

The number pulsars observed at TeV energies is still small (Table~\ref{tab:psr}), but their properties at these high energies  challenge the existing models.  
The TeV emission is best  interpreted in terms of inverse Compton up-scattering of low-energy photons (from far IR to X-rays), probably by the same population of relativistic leptons responsible for the GeV emission.
\psr , detected with HESS, is the most recent addition to this group of pulsars \citep{djannatiTP}. It is particularly interesting because its emission extends above a few TeV with a spectrally distinct hard component, a peculiar property that,  up to now, was seen only in the Vela pulsar. 
Despite the spin down power of \psr\ is  only 7\% of that of Vela, it is intrinsically more luminous, both at GeV and TeV energies, by factors $\sim$4 and $\sim$150, respectively.

The first results obtained with CTAO LST-1 for Geminga  and Crab  were also reported  \citep{yeungTP,lopez-moiaTP} and already give us  a  hint of the great progress that we can expect in this field with the complete CTAO.

Millisecond pulsars are a prominent class of sources in the MeV-GeV sky, and probably contribute a significant fraction of the sources that are still unidentified. Many of them are in binaries, as it is expected in the recycling scenario. A particularly interesting subset of millisecond binary pulsars are the so called ``spiders':  close systems in which the relativistic wind emitted by the pulsar is ablating the companion star. Although none of them has been detected yet in the VHE/UHE range, they are promising targets for CTAO, as discussed in \citet{venterTP}.

\begin{table*}
\caption{Properties of VHE pulsars}
\label{tab:psr}
\begin{center}
\begin{tabular}{lrcccccc}
\hline
Pulsar &   $P$ & $\dot{P}$ & $D$ & $\dot{E}_{\rm rot}$                            &  $\tau$  & $B_d$            &  $B_{LC}$ \\
              &[ms] &              &  [kpc] & [erg s$^{-1}$]                                     &     [kyr]   &  [G]               &   [G] \\
\hline
Crab  & 33.7 & 4.2 10$^{-13}$ &  2$^{+0.4}_{-0.3}$  & 4.3 10$^{38}$           & 1.26   &   3.8 10$^{12}$  & 9.6 10$^{5}$  \\
Vela   & 89.4 & 1.2 10$^{-13}$ &   0.28$^{+0.10}_{-0.02}$  & 6.8 10$^{36}$  & 11.3 &   3.4 10$^{12}$  & 4.4 10$^{4}$   \\
Geminga  & 237.1 & 1.1 10$^{-14}$  & 0.2$\pm$0.1  & 3.3 10$^{34}$ &           342  &   1.6 10$^{12}$   & 1.2 10$^{3}$  \\
PSR B1706--44  & 102.5 & 9.5 10$^{-14}$ &   2.6$^{+0.5}_{-0.6}$  & 3.5 10$^{36}$ &  17.5   & 3.1 10$^{12}$   & 2.7 10$^{4}$ \\
PSR J1509--5850  & 88.9 & 9.2 10$^{-15}$ &  3.4$\pm$1.3  &  5.2 10$^{35}$        &     154   &  0.9 10$^{12}$   & 1.2 10$^{4}$  \\
\hline
\end{tabular}
\end{center}
\end{table*}
 
\section{Gamma-ray binaries}

The prototype of this class   was already present among the first $\gamma$-ray sources discovered with the  COS-B satellite, although it took some time to prove that  CG 135+01  was indeed associated to the periodically radio-flaring star \lsi . This is a Be-type star in a 26.5 d orbit with a compact object, whose nature was clarified only a few years ago, thanks to the the discovery of (sporadic) radio pulsations at 269 ms with the FAST telescope \citep{2022NatAs...6..698W}. 

 About ten   $\gamma$-ray binaries  in which a compact object orbits a massive star, with periods from a few days to many years, are currently known  \citep{riboTP}.  They are called $\gamma$-ray binaries, in contrast to the much more numerous X-ray binaries, because their spectral energy distribution peaks in the HE/VHE ranges.  This reflects the different mechanism at the basis of their emission: while  X-ray binaries are powered by accretion of matter onto a collapsed star,  in the $\gamma$-ray binaries the high-energy emission is believed to originate mainly from shocks in the collisions between their powerful stellar winds.
 
  The neutron star versus  black hole nature of the compact objects in $\gamma$-ray binaries has been long debated, but the presence of a neutron star is now clear for the systems showing radio pulsations. The pulsars in these systems, contrary to the recycled millisecond pulsars mentioned in Section \ref{sec:psr}, are young and energetic ($\dot{E}_{\rm rot} > 10^{35}$  erg s$^{-1}$), which allows them to emit powerful relativistic winds.   The neutron stars in $\gamma$-ray binaries can appear as ordinary radio pulsars when they are far from periastron in wide high-eccentricity systems (e.g., PSR B1259--63). 
  On the other hand,  when they are closer to the companion star,  the radio emission is absorbed and the shocks produced by the interaction of the pulsar relativistic winds with that of the companion accelerates the particles responsible for the high-energy emission. Viewing direction effects and varying ambient conditions give rise to periodic orbital modulations of the observed flux. 
 
 These properties make
$\gamma$-ray binaries  interesting laboratories to study the same physical processes occurring in pulsar wind nebulae, but with the advantage of a time-variable physical environment and with the possibility to observe the emitting region from different directions as the neutron star moves along    its orbit.   The  bright  massive companion stars in these systems leads to the presence of a large density of photons that are important for the inverse Compton process. 

Until recently none of the  $\gamma$-ray binaries was detected at UHE, but preliminary results on the LHAASO detection of \lsi\  up to $\sim100$ TeV with a power-law spectrum have been reported at this conference \citep{liTP}.
Finally, it must be noted that high-energy emission has also been observed in massive binaries not containing collapsed stars, as demonstrated by the colliding winds binary eta Carinae \citep{2020A&A...635A.167H}.

\section{Microquasars ($\mu$QSOs)}

Microquasars, stellar mass black holes\footnote{although neutron stars cannot be excluded in some cases}  accreting from their companion stars, constitute another relevant class of galactic high-energy sources. They are found both in high-mass (e.g.  Cyg X-1, Cyg X-3) and low-mass (e.g. V4641 Sgr, V404 Cyg) binaries and, being characterised by the presence of collimated relativistic jets, are ideal laboratories to study the accretion/ejection connection on time scales more accessible than those typical of active galactic nuclei. 
Two $\mu$QSOs extensively  discussed in this conference are SS 433 and V4641 Sgr.

 SS 433 is  the most powerful $\mu$QSOs in the Galaxy: its precessing, baryon-loaded jets have a kinetic power of 10$^{39-40}$ erg s$^{-1}$ and produce X-ray hot spots and radio lobes extending up to more than 100 pc  on both sides from the central source.  The energy-dependent morphology of the VHE emission from the jets, with the higher energies closer to the central source,  locates the place of lepton acceleration at the base of the outer jets, where they emit synchrotron X-rays and inverse Compton  $\gamma$-rays \citep{2024Sci...383..402H}. Emission up to $\sim$100 TeV has been observed from the western lobe of SS 433 with HAWC \citep{2024ApJ...976...30A},
while above this energy LHAASO detected  extended emission slightly displaced from the central source \citep{2024arXiv241008988L}. The spatial coincidence with a massive molecular cloud suggests a hadronic origin for this UHE emission.

 V4641 Sgr is the second  $\mu$QSO  seen at VHE \citep{2024Natur.634..557A}. 
 The    superluminal motion observed in this source implies that its radio jets  are nearly aligned with the line of sight.
 On the other hand,  HAWC detected an  elongated source, extending tens of  parsecs in the plane of the sky, in a direction unrelated to that of the relativistic jets  \citep{2024Natur.634..557A}.  A recently proposed explanation for this misalignment is that the  $\gamma$-ray emission traces the propagation along ordered magnetic field lines of the high-energy particles accelerated in the jet \citep{2024arXiv241017608N}.
 The hard spectrum of V4641 Sgr extends up to hundreds of TeV  and suggests  a hadronic origin \citep{2024arXiv241008988L}.
 
These results and the LHAASO detection of a  few other  $\mu$QSOs \citep{2024arXiv241008988L}, as well as  the possibility that Cyg X-3 be responsible for the few photons with E$>1$ PeV seen from the Cygnus Cocoon \citep{2024SciBu..69..449L},  indicate that $\mu$QSOs should be added to the classes of sources responsible for the highest-energy  Galactic cosmic rays.

\section{Novae}

Novae,  are a   class of transient high-energy emitters discovered quite recently.  On average, one or two of them are  detected with Fermi/LAT every year \cite{fauvergeTP}. Their outbursts are caused by thermonuclear explosions on the surface of  white dwarfs accreting from binary companions, and produce powerful mass ejections. Non-thermal emission originates from particles accelerated in shocks that can form inside the ejecta or, in case of red giant companions, when the ejecta interact with their dense stellar wind.
It is not clear yet wether hadronic or leptonic processes are the main responsible for the observed high-energy emission.

 The 2021 outburst of the recurrent nova RS  Ophiuchi provided the first detection of a nova at VHE \citep{2022Sci...376...77H,2022NatAs...6..689A}. The observed emission is best explained by hadronic models. The TeV emission peaked about two days later than the GeV one.   A possible explanation in terms of     $\gamma$-ray absorption by the interaction with  optical photons emitted   during the outburst was discussed  by   \citet{phanTP}.
   
 The fact that  to date only RS Oph, a rather ordinary nova in all other respects, has been detected at VHE  is simply due to its brightness.  Therefore,  many detections  are expected in the future for other nearby recurrent novae,  with  T Coronae Borealis\footnote{T CrB is at a distance of 0.9 kpc,  wrt  2.4 kpc of RS Oph}, which is expected to undergo an outburst in 2025 \citep{2023MNRAS.524.3146S} as a primary very interesting candidate.

\section{Massive star clusters (MSCs) and superbubbles}

The relevance of young OB stars associations as  $\gamma$-ray sources and cosmic ray accelerators was recognised more than forty years ago    \citep{1979ApJ...231...95M,1983SSRv...36..173C}. In recent years they attracted increasing interest, after the detection of VHE emission from Westerlund 1 \citep{2012A&A...537A.114A,2022A&A...666A.124A}, the most massive stellar cluster in our Galaxy. 
In this conference, several talks and posters were devoted to MSCs, that were also extensively discussed as potential sources of the  Galactic cosmic rays of highest energy (e.g. \citet{peronTP}). 
An estimate of the contribution from unresolved MSCs to the galactic diffuse $\gamma$-ray emission has been presented by \citet{menchiariTP}.

At the moment there is no definite evidence in favour of a hadronic wrt leptonic origin of the  
$\gamma$-ray emission, which has now  been detected  from the direction\footnote{in some confused regions the MSC origin of the observed  $\gamma$-ray emission is unconfirmed, owing to the presence of other plausible counterparts} of about a dozen of MSCs in the Galaxy and in the Large Magellanic Cloud. 
Also the exact location of particle acceleration is still debated, with several possibilities on different spatial scales, ranging from supernovae and interactions between the winds of individual massive stars within the cluster to the whole turbulent medium in the large superbubbles surrounding the clusters. An intermediate possibility, expected to be more relevant in compact clusters than in looser associations,  places the acceleration site at the termination shock of the collective cluster wind \citep{inventarTP}. 
Magneto-hydrodynamic simulations in 3D  indicate that also in young compact clusters, the interactions between individual winds lead to magnetic field amplification (reaching B$>$100 $\mu$G) and to complex, highly heterogeneous magnetic field morphologies \citep{harerTP}.

The study of MSCs at high energies is often complicated by the fact that these objects are naturally found in crowded, star forming regions, where many other potentially emitting sources, such as young pulsars,  PWNs,  SNRs, HII regions, and massive molecular clouds are present.  
A  striking  example of the difficulties to disentangle all these possible components, is provided by the Cygnus region  \citep{yangTP,vieuTP}. The  advances expected thanks to the high angular resolution and wide field of view of the ASTRI Mini Array have also been presented \citep{daiTP,bonolloTP,peronTP}.

\section{Pulsar wind nebulae (PWNe) and TeV halos}

PWNe constitute a large population  of Galactic emitters at TeV energy, with about two dozens firmly identified (plus many candidates) among the HESS sources. At UHE energies, they are the dominant class, considering that nearly half of the sources in the  LHAASO catalog are spatially coincident with energetic PSRs, that can produce PWNe and/or TeV halos \citep{caoTP}.  The number of PWNe with confirmed TeV emission will certainly increase when better angular resolution and extensive multi-wavelength observations will disentangle confused regions. 
     
PWNe are diffuse sources emitting across the whole electromagnetic spectrum. They are formed by the interaction of the ultra-relativistic outflows from neutron stars with the surrounding medium. The Crab Nebula was the first PWN to be discovered  and it is  certainly one of the  best studied ones, but it is not representative of the whole class. 
In fact, PWNe display a variety of spectral and morphological properties that depend on their age, on the characteristics of the powering pulsar, as well as on the environment in which they evolve.
PWNs are leptonic PeVatrons, but it is unclear if they also accelerate hadrons to PeV energies. Ions extracted by the pulsars surface are greatly outnumbered by leptons due to pair cascade processes in the magnetosphere. Thus PWNs are expected to be dominated by electrons, but a small number of hadrons might still be present  or enter the nebula when it interacts with the reverse shock of the surrounding supernova  remnant.

TeV halos extending for several degrees around the nearby middle-aged pulsars Geminga and PSR B0656+14  have been discovered a few years ago with HAWC \citep{2017Sci...358..911A}. They were recognised as a new class of sources related to pulsars, but distinct from the more ubiquitous PWNs. TeV halos are produced by  inverse Compton scattering of high-energy electrons and positrons released by a PWN and diffusing in the interstellar medium. Their observed flux and radial profile imply that the diffusion coefficient is two or three orders of magnitude smaller than the average Galactic value. The reason for this difference is still unclear, but it is possibly related to the 
 presence of a large level of turbulence (and/or to a small coherence length of the turbulence itself),  either of extrinsic origin or  caused by the cosmic-rays through the excitation of streaming instabilities. 

The large angular extent and low surface brightness makes it difficult to detect and recognise TeV halos, and it is not clear if all pulsars produce them. After the discovery of the first two prototypes, the sample of TeV halo has significantly increased in number (see Table~\ref{tab:halos} for the most relevant examples), and several candidates have been recently reported \citep{wangTP}. This large sample raises the possibility that TeV halos contribute significantly to the UHE diffuse emission from the Galactic plane, as discussed, e.g., in  \citet{giacintiTP}.

\begin{table*}
\caption{Main TeV halos and candidate TeV halos around pulsars}
\label{tab:halos}
\begin{center}
\begin{tabular}{ccccccc}
\hline
\\
Source & Pulsar & $P$ &  $D$ & $\dot{E}_{\rm rot}$  &     $\tau_c$     &   References       \\
            &            &  [s] &  [kpc]       & [erg s$^{-1}$]  &    [kyr]  &  \\
  \hline
\\
 2HWC J0635+180           & Geminga                    &   0.237   & 0.250      &    3.3 10$^{34}$   &   342   &     (1)   \\
 2HWC J0700+143           &  PSR B0656+14                 &   0.385   &  0.288     &    3.8 10$^{34}$   &  111     &   (1)      \\
 LHAASO J0621+3755     &   PSR J0622+3749            &    0.333   &    1.6      &    2.7 10$^{34}$   &  208     &    (2)   \\
 LHAASO J0249+6021    &   PSR J0248+6021             &  0.217    &  2             &  2.1 10$^{35}$    & 62         & (3)   \\ 
   LHAASO J0359+5406   & PSR  J0359+5414             & 0.079     &     3.45     &  1.3 10$^{36}$    &   75       &   (4)  \\
  LHAASO J0631+1040   &  PSR J0631+1036               &  0.288    &    2.1       & 1.7 10$^{35}$     &   44.     &  (5)   \\
  LHAASO J0635+0619     & PSR J0633+0632            &   0.297     &   1.4        &  1.2 10$^{35}$    &  59     &     (6)      \\
\\
\hline
\end{tabular}
\end{center}
(1) \citet{2017Sci...358..911A},  
(2)  \citet{2021PhRvL.126x1103A},
(3) \citet{2024arXiv241004425C}, 
(4) \citet{2023ApJ...944L..29A},  
(5) \citet{2023ApJ...956...10Z},
(6) \citet{2024ApJ...968..117Z}
\end{table*}

\section{Supernova remnants }

 At this conference, supernova remnants  (SNRs) were mainly discussed in connection to the origin of cosmic rays. In fact, SNRs have been  considered since a long time as the main sources of the Galactic cosmic rays \citep{1964ocr..book.....G}.  Multi-wavelength observations clearly indicate that electrons are accelerated up to $>$TeV energies and the  $\pi^0$-decay signatures found in a few cases show that SNRs can   accelerate   to high-energies also protons.  
However, this "SNR paradigm'' for the origin of cosmic rays   is now challenged by  IACT observations showing that most SNRs, including young ones such as Cas A \citep{2024ApJ...961L..43C},  have steep spectra and/or spectral cut-offs below PeV energies.  This means that they can  contribute significantly to the galactic cosmic rays at low and medium energies, but have difficulties to explain those of the highest-energies near the spectral knee ($\sim$3 PeV).

It widely believed that very young SNRs (ages below few hundreds of years) with high shock velocities can be PeVatrons. This is difficult to test directly, but massive molecular clouds  can act as targets for energetic cosmic-rays escaping from young SNRs, thus providing evidence for ''former'' PeVatrons. 
Some of the unidentified LHAASO sources could be explained in this way  \citep{mitchellTP}.

Several talks and posters highlighted the importance of multi-wavelength studies of SNRs, both for the morphological and spectral analysis (e.g. \citep{rigoselliTP, sanoTP}. For example, the evidence for clumpy regions with magnetic field hundred times  the average value in SN 1006 supports the modelling of broad band spectra with multiple particle populations \citep{taoTP}.

\section{Large scale Galactic structures}

The Milky Way is the most prominent feature in the MeV-GeV sky, as demonstrated since the time of the SAS-2 and COS-B sky maps. Diffuse emission from the Galactic disk has been later detected also at TeV energies. The study of this  emission,  that is  produced by  the interaction of   relativistic particles with the interstellar gas and radiation field,  is extremely informative for the understanding  of the  origin, acceleration and propagation of cosmic rays \citep{revilleTP,devinTP}. 
 
Recent measurements with HAWC  (0.3-100 TeV) and   LHAASO (10-1000 TeV) indicate that the Galactic diffuse emission is  a factor $\sim$2-3 higher than that expected assuming a uniform distribution of cosmic rays interacting with the interstellar medium  \citep{2024ApJ...961..104A,2023PhRvL.131o1001C}. 
The precise  measurement of the diffuse emission at the highest energies is complicated by the systematic uncertainties related to the  subtraction of known sources. 
Although the integrated contribution from faint spatially unresolved point sources of different classes is difficult to estimate precisely, it seems that it is insufficient to explain the excess emission, especially toward the inner regions of the Galaxy. This could point to a  non homogeneous transport of the cosmic-rays through the Galaxy \citep{tjusTP,gaggeroTP}. Note that, since the most sensitive observing facilities are in the northern emisphere, the highest energy measurements do not cover the very interesting region of the Galactic center, which will be one of the main targets of CTAO South  and SWGO \citep{zaninTP,barresTP,renTP}. 

Gamma-ray observations of sky regions at high Galactic latitudes have  the potential to inform us on the cosmic rays that escape from the  disk and diffuse into the  halo. The current searches for UHE emission associated to cold gas clouds at heights of a few kpc above the Galactic disc give only weak constraints \citep{inoueTP}, but the  data of    future facilities will allow us to better assess the role of circumgalactic cosmic rays  for the Milky Way evolution. These studies will be relevant also for our understanding of large scale features such as the Fermi and eROSITA bubbles \citep{guoTP}.

\section{Conclusions}

The exciting results presented and extensively discussed at the Gamma-2024 Conference demonstrated once more the remarkable vitality of this field,  both from the observational and theoretical sides. 
As it typically occurs when a discipline evolves from the exploratory phase to maturity, the new data lead to  more questions  to be solved and, by  providing   tighter constraints,  challenge the existing theoretical models.  

Possibly the most striking outcome of the conference,  for what concerns Galactic science, is the variety of different classes of objects that should now be considered as possible sources of energetic  cosmic rays. One of the main tasks of the powerful new facilities currently under development will be to better assess the relative contributions of these classes of cosmic accelerators and help to clarify how they manage to accelerate electrons and ions to the highest observed energies.  It is also clear that a multi-wavelength approach to this problem, in synergy with other major facilities,  is essential in order to get a complete picture and discriminate between different scenarios.

\begin{acknowledgements}
All the participants contributed to make  this  conference very lively and successful and we acknowledge the LOC and SOC members for their excellent job. 
\end{acknowledgements}


\end{document}